\documentclass[12pt]{article}
\usepackage{amsmath,amsfonts,amssymb}
\usepackage{fancyhdr,graphicx,amsmath, amscd}
\usepackage{color}
\def\be{\begin{equation}}
\def\ee{\end{equation}}
\def\bea{\begin{eqnarray}}
\def\eea{\end{eqnarray}}
\def\bt{\begin{theorem}}
\def\et{\end{theorem}}
\def\bl{\begin{lemma}}
\def\el{\end{lemma}}
\def\br{\begin{remark}}
\def\er{\end{remark}}
\def\bc{\begin{corollary}}
\def\ec{\end{corollary}}
\def\bd{\begin{definition}}
\def\ed{\end{definition}}
\def\a{\alpha}
\def\g{\gamma}

\def\l{\lambda}

\def\r{\rho}

\def\G{\Gamma}
\def\D{\Delta}

\def\L{\Lambda}

\def\cB{\mathcal{B}}

\def\cD{\mathcal{D}}

\def\cF{\mathcal{F}}

\def\cH{\mathcal{H}}

\def\cK{\mathcal{K}}
\def\cL{\mathcal{L}}

\def\cS{\mathcal{S}}

\def\cU{\mathcal{U}}

\def\bbC{\mathbb{C}}

\def\bbR{\mathbb{R}}

\def\b1{B_{1}^}

\def\ba{\begin{array}}
\def\ea{\end{array}}
\def\ben{\begin{enumerate}}
\def\een{\end{enumerate}}
\newtheorem{theorem}{Theorem}
\newtheorem{lemma}{Lemma}
\newtheorem{remark}{Remark}
\newtheorem{corollary}{Corollary}

\newtheorem{definition}{Definition}
\begin{document}
\title{On the Bures Volume of Separable Quantum States
\footnote{Keywords: Bures metric, Bures volume, Separable states,
Positive partial transpose.}}
\author{Deping Ye\thanks{Department of Mathematics, Case Western Reserve University,
10900 Euclid Avenue, Cleveland OH 44106.  Email: dxy23@case.edu.}}
\date{ }
\maketitle
\begin{abstract}
We obtain two sided estimates for the Bures volume of an arbitrary
subset of the set of $N\times N$ density matrices, in terms of the
Hilbert-Schmidt volume of that subset. For general subsets, our
results are essentially optimal (for large $N$). As applications,
we derive in particular nontrivial lower and upper bounds for the
Bures volume of sets of separable states and for sets of states with
positive partial transpose.

\vskip 3mm \par PACS numbers: 02.40.Ft, 03.65.Db, 03.65.Ud,
03.67.Mn
\end{abstract}

\section{Introduction}
Quantum entanglement was discovered in 1930's \cite{EPR1,Sch1} and
is now at the heart of quantum computation and quantum
information. The key ingredients in quantum algorithms such as
Shor's algorithm for integer factorization \cite{Sh1} or
Deutsch-Jozsa algorithm (see e.g. \cite{Nie1}), are entangled
quantum states, i.e., those  states which {\it  can not} be
represented as a mixture of tensor products of states on
subsystems. Following \cite{Wer1}, states that {\it can} be so
represented are called separable states. Since determining whether
a state is entangled or separable is in general a difficult
problem \cite{Gur}, sufficient and/or necessary conditions for
separability are very important in quantum computation and quantum
information theory, and have been studied extensively in the
literature (see e.g. \cite{Ho1, Ho2, Ho3, HRP1, HRPM1, HRPM2,
Pe1}). One well-known tool is the Peres' positive partial
transpose (PPT) criterion \cite{Pe1}, that is, if a state on
$\cH=\bbC ^{D_1}\otimes \bbC ^{D_2}\cdots \otimes \bbC ^{D_n}$ is
separable then its partial transpose must be positive.
Equivalently, if a state on $\cH$ does not have positive partial
transpose, it must be entangled. This criterion works perfectly,
namely, the set of separable states $\cS=\cS(\cH)$ equals to the
set of states with positive partial transpose
$\mathcal{PPT}=\mathcal{PPT}(\cH)$ for $\cH=\bbC ^{2}\otimes \bbC
^{2}$ (two-qubits), $\cH=\bbC ^{2}\otimes \bbC ^{3}$
(qubit-qutrit), and $\cH=\bbC ^{3}\otimes \bbC ^{2}$
(qutrit-qubit) \cite{Ho1, ST1, Wor1}. However, entangled states
with positive partial transpose appear in the composite Hilbert
space $\cH=\bbC^2\otimes \bbC^4$ and $\cH=\bbC^3\otimes \bbC^3$
\cite{Ho2} (and of course in all ``larger" composite spaces; see
also \cite{Bennett1999} discussing the three-qubit case). One
striking result is that, by some measures, the positive partial
transpose criterion becomes less and less precise as $N=\prod
_{i=1}^n D_i$ grows to infinity \cite{AS1,Sz1}. This is inferred
by comparing the Hilbert-Schmidt volumes of $\cS$ and
$\mathcal{PPT}$, the estimates which rely on the special geometric
properties of the Hilbert-Schmidt metric and were obtained by
using tools of classical convexity, high dimensional probability,
and geometry of Banach spaces. The same method can also be
employed to derive tight estimates for the Hilbert-Schmidt volume
of $\cD=\cD(\cH)$ (the set of all states on $\cH$). However, a
closed expression for the exact value of this volume is known; it
was found in \cite{Zycz1} via the random matrix theory and
calculating some nontrivial multivariate integrals.

\vskip 3mm \par Compared with the Hilbert-Schmidt metric, the
Bures metric on $\cD$  \cite{Bur1, Uh0} is, in some measures, more
natural and has attracted considerable attention (see e.g.
\cite{Ditt3, Ditt2, Ditt1, Hu0, Hu1, Uh3, Uh2, Uh1}). The Bures
metric is Riemannian but not flat. It is monotone \cite{Petz1996},
i.e., it does not increase under the action of any completely
positive, trace preserving maps. It induces the Bures measure
\cite{Ben1, Hall1, Soz1}, which has singularities on the boundary
of $\cD$. The Bures volume of $\cD$ has been calculated exactly in
\cite{Soz1} and happens to be equal to the volume of an
$(N^2-1)$-dimensional hemisphere of radius $\frac{1}{2}$
\cite{Ben1, Soz1}. (This mysterious fact does not seem to have a
satisfactory explanation.) On the other hand, the precise Bures
(or Hilbert-Schmidt) volumes of $\cS$ and $\mathcal{PPT}$ are
rather difficult to calculate since the geometry of these sets is
not very well understood and the relevant integrals seem quite
intractable. These quantities can be used to measure the {\em
priori} Bures probabilities of separability and of positive
partial transpose within the set of all quantum states. (Here {\em
priori} means that the state is selected randomly according to the
Bures measure and no further information about it is available.)
For small $N$, e.g., $N=2\times 2$ and $N=2\times 3$, the Bures
volume of $\cS$ (hence of $\mathcal{PPT}$) has been extensively
studied by numerical methods in \cite{SL5, SL4, SL3, SL2, SL1,
SL0}. For large $N$, the asymptotic behavior of the
Hilbert-Schmidt volume of $\cS$ and $\mathcal{PPT}$ was
successfully studied in \cite{AS1,Sz1}. Based on that work, we
shall derive in this paper qualitatively similar ``large $N$"
results for the Bures volume. In summary, our results state that
the relative size of $\cS$ within $\cD$ is extremely small for
large $N$ (see Corollaries \ref{Bures:S:1} and \ref{Bures:S:2} for
detail). On the other hand, the corresponding relative size for
$\mathcal{PPT}$ within $\cD$
 is, in the Bures volume radius sense (see section 2 for a precise
definition), bounded from below by
a universal (independent of $N$) positive constant (see Corollary
\ref{Bures:PPT}). The conclusion is that when $N$ is large, the
{\em priori} Bures probability of finding a separable state within
$\mathcal{PPT}$ is exceedingly small. In other words, we have
shown that, as a tool to detect separability, the positive partial
transpose criterion for large $N$ {\em is not} precise in the {\em
priori} Bures probability sense. Its effectiveness to detect
entanglement is less clear (see the comments following Corollary
\ref{Bures:PPT}).

\vskip 3mm \par This paper is organized as follows. In section 2,
we review some necessary mathematical background, particularly the
background for the Hilbert-Schmidt volume and the Bures volume.
Precise statements of our main results can be found in section 3.
Section 4 explains why
our estimates are essentially optimal for general subsets of
quantum states. Section 5 contains conclusions, comments and final
remarks.

\section{Notation and Mathematical Preliminaries}
\subsection{Mathematical framework}
\vskip 2mm  We now introduce the mathematical framework and some
notation. Let $\cH$ be the (complex) Hilbert space $\bbC
^{D_1}\otimes \bbC ^{D_2}\cdots \otimes \bbC ^{D_n}$ with
(complex) dimension $N=D_1D_2 \cdots D_n$. Here we always assume
$n\geq 2$ and  $D_i\geq 2$ for all $i=1, 2, \cdots, n$. Recall
that $D_i=2$ for all $i$ corresponds to $n$-qubits, and $D_i=3$
for all $i$ corresponds to $n$-qutrits. $n=2$ corresponds to
bipartite quantum systems and $n>2$ corresponds to multipartite
quantum systems. Denote by $\cB(\cH)$ the space of linear maps on
$\cH$. Define the Hilbert-Schmidt inner product on space $\cB
(\cH)$ as $\langle A, B\rangle_{HS}= {\rm tr}(A^\dag B).$ The
subspace of $\cB(\cH)$ consisting of all self-adjoint operators is
$\cB _{sa}(\cH)$. It inherits a (real) Euclidean structure from
the scalar product $\langle \cdot ,\cdot \rangle _{HS}$ on
$\cB(\cH)$. (This is because if $A, B\in \cB _{sa}(\cH)$, then
$\langle A, B\rangle _{HS}$ must be a real number.)  $\cD$ denotes
the set of all states on $\cH$ (more precisely, states on $\cB
(\cH)$), i.e., positive (semi) definite trace one operator in
$\cB_{sa}(\cH)$:
$$\cD=\cD(\cH):=\{\rho \in \cB_{sa}(\cH), \rho \geq 0, {\rm tr} \,\rho=1\}.$$
A state in $\cD$ is said to be separable if it is a convex
combination of tensor products of $n$ states (otherwise, it is
called entangled). Denote the set of separable states by $\cS$,
then
$$\cS =\cS(\cH):={\rm conv}\{\rho _1 \otimes \cdots \otimes \rho _n, \rho _i\in
\cD (\bbC ^{D_i})\}.$$ Both $\cD$ and $\cS$ are convex subsets of
$\cB_{sa}(\cH)$ of (real) dimension $d=N^2-1$. \vskip 2mm \par
\noindent Indent: The notation $\cS(\cH)$ is in principle
ambiguous: separability of a state on $\cB (\cH)$ is not an
intrinsic property of the Hilbert space $\cH$ nor of the algebra
$\cB(\cH)$; it depends on the particular decomposition of $\cH$ as
a tensor product of (smaller) Hilbert spaces. However, this will
not be an issue here since our study focuses on fixed
decompositions.

\subsection{Hilbert-Schmidt and Bures Measures on $\cD$}
Any quantum state on $\cH$ can be represented as a density matrix,
i.e., the $N\times N$ positive (semi) definite matrix whose
diagonal elements  sum up to $1$. Therefore, any quantum state
$\r\in \cD$ has eigenvalue decomposition $\r =U\L U^\dag$ for some
unitary matrix $U\in \cU(N)$ and some diagonal matrix $\L={\rm
diag}(\l _1, \cdots, \l_N)$ with $(\l_1, \cdots, \l_N)\in \D$.
Hereafter, ${\rm Id}_N$ is the $N\times N$ identity matrix and
$U\in \cU(N)$ means that $U$ is an $N\times N$ matrix with
$UU^\dag =U^\dag U={\rm Id}_N$. We denote by $\D$ the regular
simplex in $\bbR ^N$, i.e.,
$$\D=\left\{(\l_1, \cdots, \l_N)\in \bbR^N: \l_i\geq 0, \sum _{i=1}^N
\l _i=1\right\}.$$ The Weyl chamber of $\D$ defined by the
constraint $\l_1\geq \cdots\geq  \l_N$ is denoted by $\D_1$.
Clearly, for any  $\r =U\L U^\dag$ as above and for any diagonal
matrix $B\in \cU(N)$, we have $ U \L U^\dag=U B \L B^\dag U^\dag$.
Thus, to have unique parametrization of generic states $\r=U\L
U^\dag\in \cD$, we have to restrict $(\l_1, \cdots, \l_N)$, for
instance, to $\D_1$ and select one specific point in the coset
space $\cF ^N={\cU(N)}/{[\cU(1)]^N}$  (the flag manifold).

\vskip 2mm \par We will be interested in various measures on
$\cD$. A natural restriction is to require invariance with respect
to unitary rotations. For most problems, the interesting class of
measures are those that are invariant under conjugation by a
unitary matrix. Such measures can normally be represented  as the
product of some measure on $\D_1$ and the invariant measure on
$\cF ^N$ (see \cite{Ben1, Hip1} for more on this and for the
background on the discussion that follows). The unique (up to a
multiplicative constant) invariant measure $\g$ on $\cF ^N$ is
induced by the Haar measure on the unitary group $\cU(N)$ and has
the form
$$d\g = \prod _{1\leq i<j\leq N} 2Re(U^{-1}{d}U)_{ij}Im(U^{-1}{d}U)_{ij},$$ where $U\in \cU(N)$
and ${d} U$ is the variation of $U$ such that $U+dU\in \cU(N)$.
The total $\g$ measure of $\cF ^N$ is known to be (see
\cite{Zycz1})
\begin{equation} \label{gam-measure}Z_N=\displaystyle \frac{(2\pi
)^{N(N-1)/2}}{E(N)} , \quad \mbox{where $E(N)=\prod _{j=1}^N
\Gamma(j)$}.\end{equation}  Here $\G(x)=\int
_0^{\infty}t^{x-1}e^{-t}\,dt$  is the Gamma function.

\vskip 3mm The Hilbert-Schmidt measure $V_{HS}(\cdot)$ on $\cD$,
induced by the Hilbert-Schmidt metric, may be expressed as
\cite{Zycz1}
\begin{eqnarray}\label{Hilbert} \,dV_{HS}={\sqrt{N}}\prod
_{1\leq i<j\leq N}(\lambda _i-\lambda _j)^2 \prod _{i=1}^{N-1}
\,d\lambda _i \, d\g,\end{eqnarray} where $(\lambda_1, \cdots, \l
_N)\in \D_1$. (This is just a different name for the canonical
$d$-dimensional Lebesgue measure on $\cD$.) Therefore, to obtain
the Hilbert-Schmidt volume of $\cD$, one has to calculate the
following integral \cite{Zycz1}: \begin{eqnarray}
V_{HS}(\mathcal{D})&=&\int _{\D_1\; \times \; \cF ^N}
{\sqrt{N}}\prod _{1\leq i<j\leq N}(\lambda _i-\lambda _j)^2 \prod
_{i=1}^{N-1} \,d\lambda _i \, d\g \nonumber
\\ &=&{(2\pi )^{\frac{N(N-1)}{2}} \ \sqrt{N}}\ \frac{E(N)}{\Gamma
(N^2)}\label{Hilbert-Schmidt-D }.\end{eqnarray}

\vskip 2mm \par \noindent We define $\mbox{vrad}_{HS}(\cK)$, the
Hilbert-Schmidt volume radius of $\cK\subset \cD$, to be the
radius of $d$-dimensional Euclidean ball which has the same volume
as the Hilbert-Schmidt volume of $\cK$. In other words,
$$\mbox{vrad}_{HS}(\cK)=\left(\frac{V_{HS}(\cK)}{\sigma_d
}\right)^{\frac{1}{d}},$$ where $\sigma_d  =\frac{\pi
^{{d}/{2}}}{\Gamma(1+{d}/{2})}$ is the volume of $d=N^2-1$
dimensional Euclidean ball.  For later convenience, we also denote
$\mbox{VR}_{HS}(\cK, \cL)$ as $\mbox{VR}_{HS}(\cK,
\cL)=\left(\frac{V_{HS}(\cK)}{V_{HS}(\cL)}\right)^{1/d}=\frac{\mbox{vrad}_{HS}(\cK)}{\mbox{vrad}_{HS}(\cL)}.$
It amounts to comparing the Hilbert-Schmidt volume radii of $\cK$
and $\cL$.

\vskip 3mm \par It is known that $\mbox{vrad}_{HS}(\cD)\sim
e^{-\frac{1}{4}}d^{-\frac{1}{4}}$  \cite{Sz1} by Stirling
approximation
\begin{equation}\label{Stir:1}\Gamma(z)=\sqrt{\frac{2
\pi}{z}}~{\left(\frac{z}{e}\right)}^z
\left(1+O\left(\frac{1}{z}\right)\right).\end{equation} Stirling
approximation (\ref{Stir:1}) also implies that $(\sigma_d)
^{\frac{1}{d}} \sim {\sqrt{2e \pi}}\ {d}^{-\frac{1}{2}}$ and
therefore
\begin{equation}\label{Sz-boun-3}(V_{HS}(\cD ))^{\frac{1}{d}}\sim
(4\pi^2e)^{\frac{1}{4}}d^{-\frac{3}{4}}.\end{equation} Here
$a(n)\sim b(n)$ means $\lim _{n\rightarrow \infty }a(n)/b(n) =1$.

\vskip 3mm \par An arguably more important measure in the present
context is the Bures measure (or Bures volume) $V_B(\cdot)$, which
can be written as \cite{Soz1}
\begin{eqnarray}\label{BV-1}
dV_B= \frac{2^{\frac{2-N-N^2}{2}}}{\sqrt{\l _1\cdots \l_N}} \prod
_{1\leq i<j\leq N}\frac{(\lambda_i-\lambda_j)^2}{\lambda_i
+\lambda_j}\prod_{i=1}^{N-1}\,d\lambda_i \,d\g,\end{eqnarray}
where $(\lambda_1, \cdots, \l _N)\in \D_1$. The Bures measure is
induced by the Bures distance $d_B(\cdot, \cdot)$, which may be
defined via $d_B(\varrho _1, \varrho_2)=\sqrt{2-2{\rm
tr}\sqrt{\sqrt{\varrho_1}\varrho_2\sqrt{\varrho_1}}}$ for any
states $\varrho _1, \varrho_2 \in \cD$.

\vskip 2mm \par The Bures measure has singularities (with respect
to the Hilbert-Schmidt measure) on the boundary of $\cD$. (The
boundary corresponds to {\em at least one} of the $\l_i$'s being
$0$, and if {\em two or more} of them are $0$, then some
denominators in (\ref{BV-1}) vanish.) Thanks to the work of
Sommers and Zyczkowski \cite{Soz1}, we know the precise value of
the Bures volume of $\cD$, that is
\begin{eqnarray}V_B(\cD)&=&\int _{\D_1\;\times \; \cF ^N}
\frac{2^{\frac{2-N-N^2}{2}}}{\sqrt{\l _1\cdots \l_N}} \prod
_{1\leq i<j\leq N}\frac{(\lambda_i-\lambda_j)^2}{\lambda_i
+\lambda_j}\prod_{i=1}^{N-1}\,d\lambda_i \,d\g \nonumber \\
&=&2^{1-N^2}\frac{\pi
^{{N^2}/{2}}}{\Gamma({N^2}/{2})}.\label{Bures:precise}\end{eqnarray}
As mentioned earlier, this value happens to be the $d$-dimensional
volume of the $d$-dimensional {\em hemisphere} with radius
$\frac{1}{2}$. We define $\mbox{vrad}_B(\cK)$, the Bures volume
radius of $\cK \subset \cD$, to be
$$\mbox{vrad}_B(\cK)=\left(\frac{V_B(\cK)}{\sigma_d
}\right)^{\frac{1}{d}}.$$
While comparing the Bures volume of $\cK$
with the Hilbert-Schmidt volume of the Euclidean ball does not have
immediate geometric meaning, we find this way of describing the size of
$\cK$ in the Bures volume sense convenient in our
calculations.

\vskip 3mm \par By formulas (\ref{Stir:1}) and
(\ref{Bures:precise}), one has $\mbox{vrad}_B(\cD)\sim
\frac{1}{2}$ and hence
\begin{equation}\label{Sz-boun-4}\big(V_B(\cD)\big)^{\frac{1}{d}}\sim  \sqrt{\frac{e\pi}{2}}\ \
d^{-\frac{1}{2}}.
\end{equation}

\par \noindent For later convenience, we also define the (relative) Bures volume
radii ratio of $\cK$ to $\cL$ as $\mbox{VR}_{B}(\cK,
\cL)=\left(\frac{V_{B}(\cK)}{V_{B}(\cL)}\right)^{1/d}=\frac{\mbox{vrad}_{B}(\cK)}{\mbox{vrad}_{B}(\cL)}.$
This can be used as a measure of the relative size of $\cK$ to
$\cL$ in the Bures volume sense, and clearly {\em does} have geometric meaning.

\vskip 2mm \par We refer the reader to the references
\cite{Ben1,Hall1, Hip1, Soz1, Zycz1} for more detailed background
and for motivation. In the following sections, we are interested
in the (asymptotical) behavior of $\mbox{VR}_B(\cK, \cD)$ in terms
of its relative $\mbox{VR}_{HS}(\cK, \cD)$.

\vskip 3mm \section{Main Results} \vskip 4mm

In this section, $\cK$ will be an arbitrary (Borel) subset of
$\cD$. We will estimate the Bures volume of $\cK$, in terms of the
Hilbert-Schmidt volume of $\cK$, both from below and from above.
The following lemma is our main tool to study the asymptotical
behavior of $\mbox{VR}_B(\cK, \cD)$. We point out that these
estimates are independent of the possible tensor product structure
of $\cH$.


\bl \label{Bures:Lemma:1} For any subset  $\cK$ in $\cD$ and any
$p>1$, one has
\begin{eqnarray*}
2^{\frac{N-N^2}{2}}N^{\frac{N^2-1}{2}} V_{HS}(\mathcal{K})\leq
2^{\frac{N^2+N-2}{2}}V_B(\mathcal{K})\leq
\bigg(\frac{V_{HS}(\mathcal{K})}{\sqrt{N}}\bigg)^{\frac{1}{2p}}I(p)^{\frac{2p-1}{2p}},
\end{eqnarray*} where $I(p)$ is defined as
\begin{eqnarray}\label{I(p)}
I(p):=\frac{1}{N! \ \ \Gamma \left(\frac{(p-1)N^2}{2p-1}\right)}
\bigg(\prod_{j=1}^N \frac{\Gamma\left(1+\frac{j(p-1)}{2p-1}\right)
\Gamma\left(\frac{j(p-1)}{2p-1}\right)}{\Gamma\left(\frac{3p-2}{2p-1}\right)}\bigg)
\frac{(2\pi )^{\frac{N(N-1)}{2}}}{E(N)}.\end{eqnarray} \el

\par \noindent {\bf Remark.} $I(p)$ can be defined for all $p\notin [\frac{1}{2}, 1]$
(irrespective of $N$; since the Gamma function has poles at
nonpositive integers, there are singularities in $[\frac{1}{2},1]$
whose exact locations depend on $N$.) In particular,
$I(0)=\frac{V_{HS}(\cD)}{\sqrt{N}}$. The quantity $E(N)$ was
defined in (\ref{gam-measure}).

\vskip 3mm \par \noindent {\bf Proof.} First of all, we estimate
$V_B(\cK)$ from below. To that end, define $h: \D \rightarrow
\bbR$ as $$h(\l _1, \cdots, \l_N)=\prod_{i=1}^N \lambda_i \prod
_{1\leq i<j\leq N}(\lambda_i+\lambda_j)^2.$$ Lagrange multiplier
method implies that $(1/N, \cdots, 1/N)$ is the only critical
point of $h(\l _1, \cdots, \l _N)$ in the interior of simplex
$\D$. Clearly $h(\l _1, \cdots, \l _N)$ is always $0$ on the
boundary of the simplex $\D$, which consists of sequences for
which one or more of the $\lambda_i$'s equal to $0$, and strictly
positive in the interior of the simplex $\D$. By compactness,
$h(\l _1, \cdots, \l _N)$ must have a maximum inside, and the
critical point $(1/N, \cdots, 1/N)$ must be the (only) maximizer
of $h(\l_1, \cdots, \l_N)$ on $\D$. Therefore,
\begin{equation}\label{Estimation-h}\frac{1}{\sqrt{h(\l _1, \cdots, \l _N)}}=\frac{1}{\sqrt{\l _1
\cdots \l _N}} \prod _{1\leq i<j\leq N} \frac{1}{\lambda_i
+\lambda_j}\geq
2^{\frac{N-N^2}{2}}N^{\frac{N^2}{2}}.\end{equation} By formula
(\ref{BV-1}), the Bures volume of $\cK$ equals to $\int
_{\cK}\,dV_B$, i.e.,
\begin{eqnarray*}
V_B(\mathcal{K})=\int
_{\mathcal{K}}2^{\frac{2-N-N^2}{2}}\frac{1}{\sqrt{\l _1\cdots\l
_N}} \prod _{1\leq i<j\leq N}
\frac{(\lambda_i-\lambda_j)^2}{\lambda_i +\lambda_j}
\prod_{i=1}^{N-1}\,d\lambda_i \,d\g . \end{eqnarray*} Considering
inequality (\ref{Estimation-h}) and formula (\ref{Hilbert}), one
gets
\begin{eqnarray*} V_B(\mathcal{K})&\geq & 2^{1-N^2}N^{\frac{N^2}{2}} \int _{\mathcal{K}}\prod
_{1\leq i<j\leq N}(\lambda_i-\lambda_j)^2 \prod_{i=1}^{N-1}\,d\lambda_i \,d\g\\
&=&2^{1-N^2}N^{\frac{N^2-1}{2}} V_{HS}(\mathcal{K}).
\end{eqnarray*}
\vskip 3mm  \par Next,  we will derive the upper bound, which is
more involved (and more important for our results). The subset
$\partial \D$,  the boundary of $\D$, consists of sequences for which
some $\lambda _i=0$ and
has zero $N-1$ dimensional measure. Thus, without loss of generality, we
can assume $\lambda_i>0$ for all $i=1,\cdots, N$ and,
in particular,  $\left|
\frac{\lambda_i -\lambda_j}{\lambda_i+\lambda_j}\right|<1$ for all
$i\neq j$. This implies
\begin{eqnarray}
2^{\frac{N^2+N-2}{2}}V_B(\mathcal{K})&=&\int
_{\mathcal{K}}\frac{1}{\sqrt{\l _1\cdots\l _N}} \prod _{1\leq
i<j\leq N} \frac{(\lambda_i-\lambda_j)^2}{\lambda_i +\lambda_j}
\prod_{i=1}^{N-1}\,d\lambda_i \,d\g \nonumber
\\ &<& \int _{\mathcal{K}}\frac{1}{\sqrt{\l _1\cdots\l _N}} \prod
_{1\leq i<j\leq N}|\lambda_i-\lambda_j|\prod_{i=1}^{N-1}\,d\lambda_i \,d\g  \nonumber \\
&=&\int _{\cK}f\ g \prod_{i=1}^{N-1}\,d\lambda_i \,d\g
\label{Bures:f:g}
\end{eqnarray}
where, to reduce the clutter, we denoted \begin{eqnarray*} &&g(\l
_1, \cdots, \l _N) =\frac{1}{\sqrt{\lambda_1 \cdots \lambda_N}}
\prod _{1\leq i<j\leq N}|\lambda_i-\lambda_j|^{1-\frac{1}{p}}, \\
&& f(\l _1, \cdots, \l _N)=\prod _{1\leq i<j\leq
N}|\lambda_i-\lambda_j|^{\frac{1}{p}}.\end{eqnarray*} For any
$p>\frac{1}{2}$ (so that $2p>1$), we employ the H\"{o}lder
inequality to (\ref{Bures:f:g}) and get
\begin{eqnarray}\label{Bures:f:g:1}
2^{\frac{N^2+N-2}{2}}V_B(\mathcal{K}) \leq \bigg(\int _{\cK}
f^{2p} \prod_{i=1}^{N-1}\,d\lambda_i \,d\g \bigg)^{\frac{1}{2p}}
\bigg(\int _{\cK} g^{\frac{2p}{2p-1}}\prod_{i=1}^{N-1}\,d\lambda_i
\,d\g \bigg)^{\frac{2p-1}{2p}}.
\end{eqnarray}
Substituting $f$ into the first integral of (\ref{Bures:f:g:1})
and by (\ref{Hilbert}), one has
\begin{eqnarray}\label{Hilbert:K}
\bigg(\int _{\cK} f^{2p} \prod_{i=1}^{N-1}\,d\lambda_i \,d\g
\bigg)^{\frac{1}{2p}}=\bigg(\int _{\cK} \prod _{1\leq i<j\leq N
}(\lambda_i-\lambda_j)^2 \prod_{i=1}^{N-1}\,d\lambda_i \,d\g
\bigg)^{\frac{1}{2p}}=\bigg(\frac{V_{HS}(\mathcal{K})}{\sqrt{N}}\bigg)^{\frac{1}{2p}}.
\end{eqnarray}
\par \noindent Substituting $g$ into the second integral of (\ref{Bures:f:g:1}) leads to
\begin{eqnarray}\int _{\cK} g^{\frac{2p}{2p-1}}\prod_{i=1}^{N-1}\,d\lambda_i
\,d\g &=&\int _{\cK}\prod_{1\leq i<j\leq
N}|\lambda_i-\lambda_j|^{\frac{2p-2}{2p-1}}\prod _{i=1}^N \lambda
_i ^{(\frac{p-1}{2p-1}-1)}\prod_{i=1}^{N-1}\,d\lambda_i \,d\g \nonumber \\
&\leq& \int _{\cD}\prod_{1\leq i<j\leq
N}|\lambda_i-\lambda_j|^{\frac{2p-2}{2p-1}}\prod _{i=1}^N \lambda
_i ^{(\frac{p-1}{2p-1}-1)}\prod_{i=1}^{N-1}\,d\lambda_i \,d\g ,
\label{Bures:f:g:2}
\end{eqnarray}
the inequality following just from $\cK\subset \cD$. By
(\ref{gam-measure}) and the Fubini's theorem, the last integral in
(\ref{Bures:f:g:2}) equals to
\begin{equation}\label{Bures:f:g:3}\frac{(2\pi )^{N(N-1)/2}}{E(N)} \int
_{\D_1}\prod_{1\leq i<j\leq
N}|\lambda_i-\lambda_j|^{\frac{2p-2}{2p-1}}\prod _{i=1}^N \lambda
_i ^{(\frac{p-1}{2p-1}-1)}\prod_{i=1}^{N-1}\,d\lambda_i .
\end{equation}

\vskip 2mm \par \noindent Under the condition $\frac{p-1}{2p-1}>0$
(i.e., $p>1$ or $p<1/2$), one has (see e.g. \cite{MML1, Zycz1})
$$\int _{\D}\prod_{1\leq i<j\leq N}|\lambda_i-\lambda_j|^{\frac{2p-2}{2p-1}}\prod _{i=1}^N \lambda
_i^{\frac{p-1}{2p-1}-1} \prod _{i=1}^{N-1} \,d\lambda _i
=\frac{1}{\Gamma \left(\frac{(p-1)N^2}{2p-1}\right)}
\bigg(\prod_{j=1}^N \frac{\Gamma\left(1+\frac{j(p-1)}{2p-1}\right)
\Gamma\left(\frac{j(p-1)}{2p-1}\right)}{\Gamma\left(\frac{3p-2}{2p-1}\right)}\bigg).$$
\vskip 3mm \par \noindent Taking into account that $\D$ consists
of $N!$ Weyl chambers, we conclude that the expression in
(\ref{Bures:f:g:3}) is then equal to $I(p)$. In other words, we
have shown that $$I(p)=\int _\mathcal{D}\prod_{1\leq i<j\leq N
}|\lambda_i-\lambda_j|^{\frac{2p-2}{2p-1}}\prod _{i=1}^N \lambda
_i ^{(\frac{p-1}{2p-1}-1)}\prod_{i=1}^{N-1}\,d\lambda_i \,d\g.$$
Combining this with (\ref{Bures:f:g:1}), (\ref{Hilbert:K}), and
(\ref{Bures:f:g:2}), we conclude that if $p>1$, then
$$2^{\frac{N^2+N-2}{2}}V_B(\mathcal{K})\leq
\bigg(\frac{V_{HS}(\mathcal{K})}{\sqrt{N}}\bigg)^{\frac{1}{2p}}I(p)^{\frac{2p-1}{2p}},
$$ which is the upper estimate from Lemma \ref{Bures:Lemma:1}.


\vskip 3mm \bt\label{Bures:K:0} There is a universal computable constant
$c_1>0$, such that for any Hilbert space $\cH$ and any subset
$\cK\subset \cD$,
\begin{eqnarray*}  c_1 \ \mathrm{VR}_{HS}(\cK, \cD)
\leq \mathrm{{VR}}_{B}(\cK, \cD).
\end{eqnarray*}\et

\par \noindent {\bf Proof.} Recall $d=N^2-1$. From the lower bound of Lemma
\ref{Bures:Lemma:1}, one has
$$V_B(\mathcal{K})^{\frac{1}{d}}\geq \frac{1}{2}\ (d+1)^{\frac{1}{4}}\
V_{HS}(\mathcal{K})^{\frac{1}{d}}> \frac{1}{2}\ d^{\frac{1}{4}}\
V_{HS}(\mathcal{K})^{\frac{1}{d}}.$$
Dividing both sides by
$V_B(\cD)^{\frac{1}{d}}$, one obtains
\begin{eqnarray}
\left(\frac{V_B(\mathcal{K})}{V_B(\cD)}\right)^{\frac{1}{d}}&>&
\frac{1}{2}\ d^{\frac{1}{4}}\
\left(\frac{V_{HS}(\mathcal{K})}{V_{HS}(\cD)}\right)^{\frac{1}{d}}
\left(\frac{V_{HS}({\cD})}{V_{B}(\cD)}\right)^{\frac{1}{d}}\nonumber \\
& =&\frac{1}{2}\ d^{\frac{1}{4}}
\left(\frac{\mbox{vrad}_{HS}({\cD})}{\mbox{vrad}_{B}(\cD)}\right)\
\left(\frac{V_{HS}(\mathcal{K})}{V_{HS}(\cD)}\right)^{\frac{1}{d}}.
\label{Lemma:below}
\end{eqnarray}

\par\noindent Formulas (\ref{Sz-boun-3}) and (\ref{Sz-boun-4}) imply that
$d^{\frac{1}{4}} \ \mbox{vrad}_{HS}(\cD)\sim
2e^{-\frac{1}{4}}\mbox{vrad}_B(\cD),$ i.e., $$\lim _{d\rightarrow
\infty}\frac{1}{2}\ d^{\frac{1}{4}}
\left(\frac{\mbox{vrad}_{HS}({\cD})}{\mbox{vrad}_{B}(\cD)}\right)=e^{-\frac{1}{4}}.
$$
\par\noindent Therefore, there is a (computable)
universal constant $c_1>0$, such that, for any $N$,
$$\frac{1}{2}\ d^{\frac{1}{4}} \frac{\mbox{vrad}_{HS}(\cD)}{\mbox{vrad}_B(\cD)}\geq c_1.$$
Together with (\ref{Lemma:below}),  this shows that  for any $\cK\subset
\cD$ and for any $N$, \begin{eqnarray*} \mbox{VR}_B(\cK,\cD) \geq c_1
\mbox{VR}_{HS}(\cK,\cD).
\end{eqnarray*}
{\bf Remark.} If $N$ is relatively large, the optimal constant
$c_1=c_1(N)$ is close to $e^{-1/4}$ because
$c_1(N)=\frac{d^{\frac{1}{4}}\ \mbox{vrad}_{HS}(\cD)}{2 \
\mbox{vrad}_B(\cD)}\rightarrow e^{-1/4}$ as $N\rightarrow \infty$.
For specific values of $N$, one can compute $c_1$ precisely. For
instance, $c_1(4)\thickapprox 0.7572$. Actually,
$c_1(6)\thickapprox 0.7686, c_1(8)\thickapprox 0.7728$ and
$c_1(10)\thickapprox 0.7748$ which are very close to the
$e^{-1/4}\thickapprox 0.7788$.  It appears that the sequence
$c_1(N)$ is increasing and so $c_1=c_1(4)$ should work for all
$N$, however, we do not have a rigorous proof.


\vskip 3mm \bt \label{Bures:K:2} There is a universal computable constant
$C_1>0$ such that, for any Hilbert space $\cH$ and any $\cK\subset
\cD$,
\begin{eqnarray*}
\mathrm{VR}_B(\cK, \cD) \leq C_1 \ \sqrt{\a} \ \exp\left(\frac{\ln
\ln (e/ \a)}{2N}\right)\end{eqnarray*} where
$\a=\mathrm{VR}_{HS}(\cK, \cD)$.\et

\vskip 3mm \par\noindent {\bf Proof.} Recall $N=D_1D_2 \cdots D_n$
and $d=N^2-1$. For any $p>1$, Lemma
 \ref{Bures:Lemma:1} implies that
\begin{eqnarray*}
2^{\frac{d}{2}}V_B(\mathcal{K}) &\leq&
2^{\frac{N^2+N-2}{2}}V_B(\mathcal{K})\leq \a ^{\frac{d}{2p}}\
(V_{HS}(\mathcal{D}))^{\frac{1}{2p}}\ I(p)^{\frac{2p-1}{2p}}.
\end{eqnarray*}
Let $\beta=\frac{p-1}{2p-1}$. Replacing $V_{HS}(\cD)$ and $I(p)$
by formula (\ref{Hilbert-Schmidt-D }) and formula (\ref{I(p)}),
one has
\begin{eqnarray*}
 2^{\frac{d}{2}}V_B(\mathcal{K})&\leq &
\a^{\frac{d}{2p}}\ (2\pi)^{\frac{N^2-N}{2}} \
\frac{[E(N)]^{\frac{1}{p}-1} \
N^{\frac{1}{4p}}}{[N!]^{1-\frac{1}{2p}}}\ \ \frac{[\prod_{j=1}^N
\big(\G(\beta j) \G(1+\beta j) \big)]^{1-\frac{1}{2p}}}{[\G
(N^2)]^{\frac{1}{2p}}\ [\G (\beta N^2)
\G(1+\beta)^N]^{1-\frac{1}{2p}}}.\end{eqnarray*} Clearly
$[E(N)]^{\frac{1}{p}-1}\leq 1$ and
${N^{\frac{1}{4p}}}{(N!)^{\frac{1}{2p}-1}}\leq 1$ if $p>1$. Hence,
\begin{eqnarray}\label{bound:4:10} V_B(\mathcal{K}) &\leq & \a^{\frac{d}{2p}}\
\pi^{\frac{d}{2}}\  \frac{[\prod_{j=1}^N \big(\G(\beta j)
\G(1+\beta j) \big)]^{1-\frac{1}{2p}}}{[\G (N^2)]^{\frac{1}{2p}}\
[\G (\beta N^2) \G(1+\beta)^N]^{1-\frac{1}{2p}}}.
\end{eqnarray}

\vskip 3mm \par Since $x \Gamma(x) = \Gamma(x+1)$, it is easy to
see that for all $x\in (0,1)$ the upper estimate of $\G (x)$ is
$\frac{1}{x}$ and (somewhat less easy that) the lower estimate of
$\G(x)$ is $\frac{1}{\vartheta x}$, where $\vartheta\thickapprox
1.12917$ \cite{Gammamin}. That is
\begin{equation}\label{YD3} \frac{1}{\vartheta x}\leq \G(x)\leq
\frac{1}{x},\ \ \mbox{or} \  \ \frac{1}{\vartheta }\leq
\G(1+x)\leq 1, \ \ \ \mbox{for all $x\in (0,1)$}.\end{equation}
 \par \noindent Pick
$p=p(N,\a):=\frac{N^2\ln(e/\a)-1}{N^2\ln(e/\a)-2}$ as a function
of $N$ and $\a$, so that $\beta =\frac{p-1}{2p-1} =\frac{1}{N^2\ln
(e/\a)}$. Equivalently, $\beta N^2 =\frac{1}{\ln (e/\a)}$. Since
$\a\leq 1$ for all $\cK\subset \cD$, then $\beta N^2 <1$ and hence
$\beta j\leq 1$ for all $j=1, 2, \cdots N^2$.
Taking inequality
(\ref{YD3}) into account, one has
\begin{eqnarray}\label{YD39}
\left[\frac{\G(1+\beta j)}{\G(1+\beta)}\right]^{1-\frac{1}{2p}}
\leq \vartheta, \quad \mbox{for all $j=1, 2, \cdots N$.}
\end{eqnarray}
Consequently, again by inequality (\ref{YD3}), and $N\geq 4$,
\begin{eqnarray}\label{YD40}
& & \left[\frac{\prod_{j=1}^N \G(\beta j)}{\G (\beta N^2)
}\right]^{1-\frac{1}{2p}} \leq \left(\frac{\vartheta \
N}{(N-1)!}\right)^{{1-\frac{1}{2p}}}\ \beta
^{(1-N)(1-\frac{1}{2p})} \leq \beta ^{(1-N)(1-\frac{1}{2p})}.
\end{eqnarray}
Combining inequality (\ref{bound:4:10}) with inequalities
(\ref{YD39}) and (\ref{YD40}), one has
\begin{eqnarray*}
V_B(\mathcal{K}) &\leq & \a^{\frac{d}{2p}}\ \pi^{\frac{d}{2}}\
\vartheta^{N}\ \beta ^{(1-N)(1-\frac{1}{2p})}\ {[\G
(N^2)]^{\frac{-1}{2p}}}.
\end{eqnarray*}
Equivalently, taking $d$-th root from both sides,
\begin{eqnarray*}
(V_B(\mathcal{K}))^{\frac{1}{d}} &\leq & \a^{\frac{1}{2p}}\
\pi^{\frac{1}{2}}\ \vartheta^{\frac{1}{N-1}}\ \beta
^{\frac{-1}{N+1}(1-\frac{1}{2p})}\ {[\G (N^2)]^{\frac{-1}{2pd}}}.
\end{eqnarray*}
Note $N\geq 4$, and hence $\vartheta^{\frac{1}{N-1}}\leq \vartheta
^{\frac{1}{3}}=1.0413$. Now dividing $V_B(\cD)^{\frac{1}{d}}$ from
both sides of the above inequality, one gets
\begin{eqnarray}\label{YD4-0}\mbox{VR}_B(\cK, \cD) &\leq & 2\vartheta
^{\frac{1}{3}} \ \a^{\frac{1}{2p}} \ [N^2\ln
(e/\a)]^{\frac{1}{N+1} (1-\frac{1}{2p})}\
[\G\left({N^2}/{2}\right)]^{\frac{1}{d}}\ {[\G
(N^2)]^{\frac{-1}{2dp}}}.\end{eqnarray} It is easy to verify that
\begin{eqnarray*}&&[N^2\ln (e/\a)]^{\frac{1}{2(N+1)}} \leq
\exp\left(\frac{\ln N}{N}\right)\ \exp\left(\frac{\ln \ln
(e/\a)}{2N}\right)\leq \sqrt{2} \ \exp\left(\frac{\ln \ln
(e/\a)}{2N}\right),\\
&& [N^2\ln (e/\a)]^{\frac{1}{2(N+1)}
(1-\frac{1}{p})}=\exp\left(\frac{\ln(N^2 \ \ln (e/\a))
}{2(N+1)(N^2\ln(e/\a)-1)}\right)\leq
\exp\left(\frac{1}{N}\right)\leq e^{\frac{1}{4}}.\end{eqnarray*}
Therefore,
\begin{eqnarray}\label{YD4-1} [N^2\ln
(e/\a)]^{\frac{1}{N+1} (1-\frac{1}{2p})} \leq \sqrt{2}\
e^{\frac{1}{4}} \ \exp\left(\frac{\ln \ln
(e/\a)}{2N}\right)\end{eqnarray} Also, we can verify that
\begin{eqnarray}\label{YD4} && \a ^{\frac{1}{2p}}=\a ^{\frac{1}{2}}\ \exp\left(\frac{\ln
(1/\a)}{2[N^2(1+\ln (1/\a))-1]}\right)\leq \a^{\frac{1}{2}}\
\exp\left(\frac{1}{2N^2}\right)\leq e ^{\frac{1}{32}} \ \sqrt{\a
}.\end{eqnarray} Since $\G(N^2)\leq (N^2)!\leq \exp(N^2\ln N^2)$,
one has
\begin{eqnarray}\label{YD4-2}
\G(N^2)^{\frac{1}{2d}-\frac{1}{2pd}}&=&\exp\left(\frac{\ln
(\G(N^2))}{2(N^2-1)\ (N^2\ln(e/\a)-1)}\right)\nonumber \\ &\leq&
\exp\left(\frac{2 \ln N}{N^2\ln(e/\a)}\right)\leq
\exp\left(\frac{2}{N}\right)\leq e^{\frac{1}{2}}.\end{eqnarray}
Stirling approximation formula (\ref{Stir:1}) implies that
\begin{equation}\label{Gammaratio}\lim _{N\rightarrow \infty}
\frac{\G(N^2/2)^{\frac{1}{d}}}{[\G
(N^2)]^{\frac{1}{2d}}}=\frac{1}{\sqrt{2}}.\end{equation} Together
with inequalities (\ref{YD4-0}), (\ref{YD4-1}),(\ref{YD4}), and
(\ref{YD4-2}), there exists a universal (independent of $N, \a$)
constant $C_1>0$, such that, $\mbox{VR}_B(\cK, \cD) \leq  C_1
\sqrt{\a} \ \exp\left(\frac{\ln\ln(e/\a)}{2N}\right).$

\vskip 2mm \par \noindent {\bf Remark.} A slightly more precise
calculation shows that
\begin{eqnarray*}
\mbox{VR}_B(\cK, \cD) \leq \sqrt{2 \a} \ \exp\left(\frac{\ln \ln
(e/ \a)}{2N}\right) \ \left[1+O\left(\frac{\ln
N}{N}\right)\right].\end{eqnarray*} The calculation yields
explicit (not necessarily optimal) values of $C_1$ in the theorem.
For small dimensions, our proof yields $C_1(4)\thickapprox 2.5164$
if $N=4$, $C_1(6)\thickapprox 2.2137 $, and $C_1(8)\thickapprox
2.0478$. As the dimension $N$ becomes large, the value of  $C_1$
given by the argument tends to $\sqrt{2}\thickapprox 1.4142$. On
the other hand, the Legendre duplication formula (see
\cite{Abramowitz1972}) says that
$$\Gamma(z) \; \Gamma\left(z + {1}/{2}\right) = 2^{1-2z} \; \sqrt{\pi} \;
\Gamma(2z).$$ By taking $z={N^2}/{2}$, one can rewrite the
expression in (\ref{Gammaratio}) as
\begin{equation}\label{Gammaratio1}\left(\frac{\G(N^2/2)\ \G(N^2/2)}{\G
(N^2)}\right)^{\frac{1}{2(N^2-1)}}=\frac{1}{\sqrt{2}}\
\left(\frac{\sqrt{\pi}\ \G
(N^2/2)}{\G({N^2/2+1/2})}\right)^{\frac{1}{2(N^2-1)}}.\end{equation}
Gamma function is log-convex \cite{Mollerup1922}, and hence
$$ \Gamma({{N^2}/ 2})^2 \leq \Gamma({{N^2}/2}-{1/2})
\Gamma({{N^2}/2}+{1/2}) =\frac{\Gamma({{N^2}/2}+{1/2}) ^2}{{{N^2}/
2}-{1/2}}.$$ Equivalently
$$\frac{\Gamma({{N^2}/2}+{1/2})}{\Gamma({{N^2}/2})}\geq
\sqrt{{\frac{N^2-1}{2}}},$$ which is greater than $\sqrt{\pi}$ iff
$N > \sqrt{2\pi+1} \approx 2.7$. Together with formula
(\ref{Gammaratio1}), this shows that the asymptotic relation
(\ref{Gammaratio}) is in fact an upper bound for all $N\geq 3$. It
follows that $C_1\thickapprox 2.5164$ works for all $N\geq 4$,
$C_1\thickapprox 2.2137$ works for all $N\geq 6$, etc.

\vskip 2mm \par \noindent {\bf Remark.} In most cases of interest
$\a$ is such that the factor $\exp\left(\frac{\ln \ln (e/
\a)}{2N}\right)$ is bounded by a universal numerical constant.
 For instance, if  $\ln (1/\a) \leq a_1 \ e^{a_2 N}$ for
some constants $a_1>0, a_2>0$, then
$$\mbox{VR}_B(\cK,\cD) \leq \sqrt{2 e^{a_2}}\
\sqrt{\mbox{VR}_{HS}(\cK, \cD)} \ \left[1+O\left(\frac{\ln
N}{N}\right)\right].$$ While our argument doesn't give similar
estimates for general $\a$, other ways of writing the estimates in
more transparent ways are possible. For example,  for any fixed
$p>1$ there is a constant $C_p>0$ depending on $p$ (but
independent of $N$ and $\a$), such that
$$\mbox{VR}_B(\cK, \cD) \leq C_p \ \big(\mbox{VR}_{HS}(\cK,
\cD)\big)^{\frac{1}{2p}}.$$

\vskip 3mm \par \noindent In the cases of $\cS$ and
$\mathcal{PPT}$, $\frac{1}{\a}$ is bounded from above by $N^k$ for
some (fixed) integer $k$ \cite{AS1, Sz1}. Therefore,
$\mbox{VR}_B(\cS, \cD) \leq \tilde{C}_1 \
\sqrt{\mbox{VR}_{HS}(\cS, \cD)}$ where $\tilde{C}_1>0$ is a
universal constant independent of $N$. Similarly,
$\mbox{VR}_B(\mathcal{PPT}, \cD) \leq \bar{C}_1 \
\sqrt{\mbox{VR}_{HS}(\mathcal{PPT}, \cD)}$ where ${\bar{C}}_1>0$
is a universal constant independent of $N$. \vskip 2mm \par
\noindent  {\bf Remark.} We point out that there is a lot of
flexibility in the choice of $\beta=\frac{1}{N^2\ \ln (e/\a)}$
(hence the choice of $p(N,\a)$). For example, one can choose
$\beta=\frac{1}{e^N\ \ln (e/\a)}$, and proves Theorem
\ref{Bures:K:2} with different (larger) constants. However,
formula (\ref{YD4}) does suggest that the factor $\ln(e/\a)$ in
$\beta$ is essentially optimal in general.

\vskip 3mm \par As applications of Theorems \ref{Bures:K:0} and
\ref{Bures:K:2}, and the estimates for $\mbox{VR}_{HS}(\cS, \cD)$
implicit in \cite{AS1}, one immediately has the following
corollaries. \bc \label{Bures:S:1}{\bf (Large number of small
subsystems)} For system $\cH=(\bbC^D)^{\otimes n}$, there exist
universal computable constants $c_2, C_2>0$, such that for all
$D,n\geq 2$,
\begin{equation*} \label{YD1} \frac{c_2}{N^{{1}/{2}+\a _D}} \leq
\mathrm{VR}_{B}(\cS, \cD) \leq C_2 \ \sqrt{\frac{(D n\ln
n)^{1/2}}{N^{1/2+\a _D}}},
\end{equation*}
where $\a _D =\frac{1}{2}\log _D
(1+\frac{1}{D})-\frac{1}{2D^2}\log _D(D+1)$.\ec

\vskip 3mm \bc \label{Bures:S:2}{\bf (Small number of large
subsystems)} For system $\cH=(\bbC^D)^{\otimes n}$, there exist
universal computable constants $c_3, C_3>0$, such that for all $D,n\geq 2$,
\begin{equation*} \label{YD2} \frac{c_3^n}{N^{1/2-1/(2n)}} \leq
\mathrm{VR}_{B}(\cS, \cD) \leq C_3 \ \sqrt{\frac{(n \ln
n)^{1/2}}{N^{1/2-1/(2n)}}}. \end{equation*} \ec {\bf Remark.}
Recall that if $\cH=(\bbC^D)^{\otimes n}$, then the dimension of
$\cH$ is $N=D^n$ and so the expressions in the the numerators of
the estimates in the Corollaries above are of smaller order than
the denominators. Hence, for any fixed small $D$, Corollary
\ref{Bures:S:1} shows that $\mbox{VR}_{B}(\cS, \cD)$ goes to $0$
exponentially as $n\rightarrow \infty$. On the other hand, for
$\cH=(\bbC^D)^{\otimes n}$ and for fixed small $n$, Corollary
\ref{Bures:S:2} shows that ``the order of decay" of $VR_B(\cS,
\cD)$ is between $D^{\frac{1}{2}-\frac{n}{2}}$ and
$D^{\frac{1}{4}-\frac{n}{4}}$ as $D\rightarrow \infty$. In both
cases, the {\em priori} Bures probability of separability is
extremely small for large (and even for moderate) $N$. It is
possible to provide (not necessarily optimal) estimates on the
constants appearing in both corollaries. For instance, in
Corollary \ref{Bures:S:1}, one can take $c_2(4)=0.2272,
C_2(4)=\sqrt{4.4}\ C_1(4)=5.2785$ if $N=4$, $c_2(6)=0.2306,
C_2(6)=\sqrt{4.4}\ C_1(6)=4.6436$, and $c_2(8)= 0.2318,
C_2(8)=\sqrt{4.4}\ C_1(8)= 4.2955$. If the dimension $N$ is large
(particularly for large $n$), the relevant asymptotic behaviors of
$c_2$ and $C_2$ given by the proofs are: $c_2$ tends to
$\sqrt{\frac{e}{2\pi}}\thickapprox 0.6577$, and $C_2$ tends to
$2^{3/4}e^{1/8}\thickapprox   1.9057$. In Corollary
\ref{Bures:S:2}, $c_3$ can be taken as
$c_3=e^{-1/4}/\sqrt{6}\thickapprox  0.3179$ and $C_3=C_2$. We
refer the readers to \cite{AS1} for the constants in terms of the
Hilbert-Schmidt volume.


\vskip 3mm \par In the rest of this section, we will discuss the
Bures volume of $\mathcal{PPT}$. For a bipartite system $\cH=\bbC
^{D_1} \otimes \bbC ^{D_2}$, any state $\rho $ on $\cH$ can be
expressed uniquely as
$$\rho = \sum _{i,j}^{D_1} \sum _{\a, \beta }^{D_2} \rho _{i\a, j\beta } |e_i\otimes
f_{\a}\rangle \langle e_j\otimes f_{\beta}|$$ where
$\{e_i\}_{i=1}^{D_1}$ and $\{f_\a\}_{\a=1}^{D_2}$ are the
canonical bases of $\bbC ^{D_1}$ and $\bbC ^{D_2}$ respectively.
Define the partial transpose $T(\rho)$ with respect to the first
subsystem as
$$T(\r)=\sum _{i,j}^{D_1} \sum _{\a, \beta }^{D_2} \rho _{j\a, i\beta } |e_i\otimes
f_{\a}\rangle \langle e_j\otimes f_{\beta}|.$$ We write
$\mathcal{PPT}$ for the set of states $\r$ such that $T(\r)$ is
also positive. (Note that $\mathcal{PPT}$ is basis-independent
because eigenvalues do not depend on a basis \cite{Pe1}.) The
Peres criterion asserts: {\em every separable state has a positive
partial transpose} \cite{Pe1}. That is, $\cS \subset \mathcal{PPT}
\subset \cD$. For qubit-qubit system $\bbC^2\otimes \bbC^2$ and
qubit-qutrit (or qutrit-qubit) system $\bbC^2\otimes \bbC^3$ (or
$\bbC^3\otimes \bbC^2$), the positive partial transpose criterion
gives a sufficient and necessary condition for separability, i.e.,
$\mathcal{S}=\mathcal{PPT}$ \cite{Ho1, ST1, Wor1}. The following
corollary, which is a direct consequence of Theorems
\ref{Bures:K:0}, \ref{Bures:K:2}, and of Theorem 4 of \cite{AS1},
gives the estimation of the Bures volume of $\mathcal{PPT}$.

\bc\label{Bures:PPT} {\bf (Bures Volume of $\mathcal{PPT}$):}
There exists an absolute computable constant $c_0>0$, such that, for any
bipartite system $\cH=\bbC ^D\otimes \bbC ^D$, $c_0\leq
\mathrm{VR}_B(\mathcal{PPT}, \cD)\leq 1.$ \ec

\par \noindent Indent: An unsolved question is: {\em does there exist a universal constant
$0<C_0<1$ such that $\mathrm{VR}_B(\mathcal{PPT}, \cD)\leq
C_0<1$?} Answering this question would help us understand the
effectiveness of positive partial transpose criterion as a tool to
detect quantum entanglement for all $D \geq 3$.
(The answer to the analogous question about $\mathrm{VR}_{HS}$
 is not known, either.)

\vskip 3mm \par An immediate consequence of Corollaries
\ref{Bures:S:2} and \ref{Bures:PPT} is that, for $\cH=\bbC
^D\otimes \bbC ^D$ and large $D$, there exist universal constants
$c_4, C_4>0$ (independent of $D$), such that, $c_4
{D^{-\frac{1}{2}}} \leq \mbox{VR}_{B}(\cS, \mathcal{PPT}) \leq
{C_4} {D^{-\frac{1}{4}}}$. The upper bound decreases to $0$
 as $D\rightarrow \infty$. In other word, the conditional
{\em priori} Bures probability of separability given positive
partial transpose condition is exceedingly small. Hence,
for large $N$,  the PPT
criterion is not precise as a tool to detect separability.


\section{Optimality of the bounds}
In this section, we will prove that, in general, the bounds in
Theorems \ref{Bures:K:0} and \ref{Bures:K:2} are essentially
optimal. The Bures volume has singularities close to the boundary
of $\cD$, so the optimal upper bound is intuitively attained by
the subsets close to the boundary of $\cD$. On the other hand, for
the subsets located near the maximal state $\r _{max}=\frac{{\rm
Id}_N}{N}$, we can achieve the lower bound (this is really a
simple consequence of the proof of Lemma \ref{Bures:Lemma:1}).

\vskip 3mm
\subsection{Optimality of the lower bound}
\vskip 3mm For $0<t<1$, let $\cK_t=t \cD +(1-t )\r _{max}$, i.e.,
$$\cK_t=\left\{UXU^\dag: X={\rm diag}\left(\frac{1-t}{N}+t\lambda_1,
\cdots, \frac{1-t}{N}+t\lambda_N\right),\ \mbox{$(\l _1,\cdots
\l_N)\in \D$ and $U\in \cU(N)$} \right\}.$$  Let $Z_N$ be as in
(\ref{gam-measure}).

\vskip 3mm \par We now estimate $V_B(\cK_t)$ from above. By
formula (\ref{BV-1}),
\begin{eqnarray*} 2^{\frac{N^2+N-2}{2}}\ V_B(\cK_t)&=&  \int _{\D_1} \frac{Z_N \
t^{N^2-1}}{\sqrt{\prod _{i=1}^N (t\l _i +\frac{1-t}{N})}}\prod
_{1\leq i<j\leq N} \frac{(\l _i -\l _j)^2}{(\frac{2-2t}{N}+t\l_i
+t\l_j)} \, \prod _{i=1}^{N-1}\,d\l_i.
\end{eqnarray*}
As $\frac{1-t}{N}+t\l_i \geq \frac{1-t}{N}$ for all $i$, one
obtains
\begin{eqnarray*}2^{\frac{N^2+N-2}{2}}\ V_B(\cK_t) &\leq & Z_N \ \frac{t^{N^2-1}\ N^{N^2/2}}{2^{(N^2-N)/2}\
(1-t)^{N^2/2}}\int _{\D_1} \prod _{1\leq i<j\leq N} (\l _i -\l
_j)^2\, \prod _{i=1}^{N-1}\,d\l_i\\ &=& \frac{t^{N^2-1}\
N^{(N^2-1)/2}}{2^{(N^2-N)/2}\ (1-t)^{N^2/2}} \
V_{HS}(\cD),\end{eqnarray*} where the equality follows the formula
(\ref{Hilbert}). By Lemma \ref{Bures:Lemma:1}, one gets
\begin{eqnarray*} V_B(\cK_t) \leq \frac{t^{N^2-1}}{(1-t)^{\frac{N^2}{2}}} \ V_B(\cD)
\leq \frac{t^{N^2-1}}{(1-t)^{N^2-1}} \ V_B(\cD).\end{eqnarray*}
Hence, $\mbox{VR}_B(\cK_t, \cD)\leq \frac{t}{1-t}\leq 4t$ for all
$t\leq \frac{3}{4}$. On the other hand, $\mbox{VR}_{HS}(\cK_t,
\cD)=t$ holds trivially because of the homogeneity of the
Hilbert-Schmidt measure.  We have proved that $\mbox{VR}_B(\cK_t,
\cD)\leq 4\mbox{VR}_{HS}(\cK_t, \cD)$ for all $\cK _t$ such that
$\mbox{VR}_{HS}(\cK_t, \cD)\leq \frac{3}{4}$.

\vskip 3mm \par Theorem \ref{Bures:K:0} guarantees that the lower
bound of $\mbox{VR}_B(\cK_t, \cD)$ is at least (up to a
multiplicative constant) $\mbox{VR}_{HS}(\cK_t, \cD)$. So the
lower bound of $\mbox{VR}_B(\cK_t, \cD)$ in Theorem
\ref{Bures:K:0} can be obtained, and hence is optimal in general.

\subsection{Optimality of the upper bound}
For $0<t<1$, we consider $\cK^t$ as
$$\cK^t=\{UXU^\dag: X={\rm diag}(1-t+t\lambda_1, t\lambda_2, \cdots,
t\lambda_N), \ \mbox{$(\l _1,\cdots \l_N)\in \D_1$ and $U\in
\cU(N)$} \}.$$ Recall $\D _1$ is the chamber of $\D$ with order
$\l_1 \geq \cdots \geq \l _N$.

\vskip 3mm \par The Hilbert-Schmidt volume of $\cK ^t$ can be
calculated by the following integral
$$V_{HS}(\cK^t)= Z_N \ t^{(N-1)^2}\ \sqrt{N}\int _{\D_1} \prod _{2\leq i<j\leq N}(\l _i -\l _j)^2 \
\prod _{k=2}^N (t\l _1 -t\l _k+1-t)^2  \, \prod
_{i=1}^{N-1}\,d\l_i.$$ As $0\leq t\l _1 -t\l _k+1-t \leq 1$, one
has
\begin{eqnarray}
V_{HS}(\cK^t)&\leq & Z_N\ t^{(N-1)^2}\sqrt{N}\int _{\D_1} \prod
_{2\leq i<j\leq N}(\l _i -\l _j)^2 \, \prod _{i=1}^{N-1}\,d\l_i \nonumber \\
&=&  \frac{Z_N}{Z_{N-1}}\ t^{(N-1)^2}\ \sqrt{\frac{N}{N-1}}\
V_{HS}(\cD _{N-1}) \ \int _{0}^1 (1-\l_1)^{N^2-2N}\,d\l_1
\nonumber\\&\leq& \sqrt{2}\
 \frac{Z_N}{Z_{N-1}}\ t^{(N-1)^2}\
V_{HS}(\cD _{N-1}) \label{Upp:1}.
\end{eqnarray}
By Stirling approximation (\ref{Stir:1}), $ \mbox{VR}_{HS}(\cK^t,
\cD) \leq C_4\ t \ t^{\frac{-2}{N+1}}$ holds for some universal
constant $C_4>0$. If $t^{\frac{-2}{N+1}}$ is bounded from above,
e.g., $t>e^{c_4 (-1-N)}$ for some constant $c_4>0$, then
\begin{equation}\label{VRHS:K:D} \mbox{VR}_{HS}(\cK ^t , \cD)\leq
C_5\ t \end{equation} holds for a new universal constant $C_5>0$.

\vskip 3mm \par  Next, we estimate the Bures volume of $\cK^t$ from
below. By formula (\ref{BV-1}),
\begin{eqnarray*} && 2^{\frac{N^2+N-2}{2}}\ V_B(\cK^t)\\& &=
\int _{\D_1} \frac{Z_N\ t^{\frac{(N-1)^2}{2}}}{\sqrt{(t\l _1 +1-t)
\l_2\cdots \l _N }}\prod _{2\leq i<j\leq N}\frac{(\l _i -\l
_j)^2}{\l_i +\l_j} \ \prod _{k=2}^N \frac{(t\l _1 -t\l
_k+1-t)^2}{t\l _1 +t\l _k+1-t} \, \prod _{i=1}^{N-1}\,d\l_i \\
& &\geq  Z_N\ \int _{\D_1} \frac{(1-t)^{2N-2}\
t^{\frac{(N-1)^2}{2}}}{\sqrt{\l_2\cdots \l _N }}\prod _{2\leq
i<j\leq N}\frac{(\l _i -\l _j)^2}{\l_i +\l_j} \, \prod
_{i=1}^{N-1}\,d\l_i
\end{eqnarray*}
where the inequality is because of $0\leq t\l _1 +1-t\leq 1$,
$0\leq t\l _1 +t\l _k+1-t\leq 1$ and $t\l _1 -t\l _k+1-t\geq 1-t$.
The last integral can be computed as in (\ref{Upp:1}) and leads to
\begin{eqnarray*}
&& 2^{\frac{N^2+N-2}{2}}\ V_B(\cK^t)\\
& & \geq
\frac{Z_N}{Z_{N-1}}\ (1-t)^{2N-2}\ t^{\frac{(N-1)^2}{2}}\
2^{\frac{N^2-N-2}{2}}\ V_{B}(\cD_{N-1})\ \int _{0}^1
(1-\l_1)^{\frac{N^2-2N-1}{2}} \,d\l_1
\\ & & =\frac{Z_N}{Z_{N-1}}\  \frac{2^{\frac{N^2-N}{2}}\ (1-t)^{2N-2}\ t^{\frac{(N-1)^2}{2}}}
{(N-1)^2}\ V_{B}(\cD_{N-1}).\end{eqnarray*} Employing the Stirling
approximation (\ref{Stir:1}) one gets $\mbox{VR}_B(\cK ^t,
\cD)\geq c_5 \ \sqrt{ t}$ for some universal constant $c_5>0$ if,
say, $t<\frac{4}{5}$.  Together with (\ref{VRHS:K:D}), we have
thus proved $$\mbox{VR}_B(\cK ^t, \cD) \geq \bar{c}_5
\sqrt{\mbox{VR}_{HS}(\cK ^t, \cD)}$$ for some universal constant
$\bar{c}_5>0$, if $t<\frac{4}{5}$ and  $t>e^{c_4 (-1-N)}$. Theorem
\ref{Bures:K:2} guarantees that $\mbox{VR}_B(\cK ^t, \cD) \leq C_1
\ \sqrt{\mbox{VR}_{HS}(\cK ^t, \cD)}.$ Therefore, the upper bound
in Theorem \ref{Bures:K:2} can also be achieved, and is optimal in
general.

\section{Conclusion and Comments}

In summary, we proved that
if $\cK$ is a Borel subset of $\cD$, then the {\em priori} Bures probability
of $\cK$ can be estimated from above and from below in terms of
the {\em priori} Hilbert-Schmidt probability of $\cK$.
Specifically, under some mild conditions on $\cK$
the relative Bures volume radius  $\mbox{VR}_{B}(\cK, \cD)$ can be
(approximately)
bounded  from below by the relative Hilbert-Schmidt volume radius
$\mbox{VR}_{HS}(\cK, \cD)$,  and from above by $\sqrt{\mbox{VR}_{HS}(\cK, \cD)}$.
We employ these results to estimate the Bures volume of $\cS$
and $\mathcal{PPT}$ and the relevant {\em priori} Bures probabilities.
 We deduce that positive partial transpose
criterion becomes less and less precise as the dimension of $\cH$
becomes larger and larger, at least if the goal is to detect
separability. We also give examples showing that, for general
subsets, our bounds are essentially optimal.

\vskip 1mm When $N$ is small, for instance when $N=4$ or $6$, our
estimates for $V_B(\cS)$ are less precise than Slater's numerical
results. However, our methods overcome the big disadvantage of
the numerical approach,  which works only for small
$N$. Moreover, our results are independent of the structure of
$\cH$. (In applications, of course, the information on the structure of
$\cH$ will be hidden in the calculation of $\mbox{VR}_{HS}(\cK,
\cD)$.)  Proceeding along similar lines one can obtain similar results
for real Hilbert spaces, and then estimate the Bures volume
of $\cS$ or $\mathcal{PPT}$ on a real Hilbert space.

\vskip 1mm  As is well known, for sets in a Euclidean space
(in particular, for sets of matrices endowed with the  Hilbert-Schmidt
metric) the volumetric  information is roughly equivalent to the
metric entropy information such as covering and packing numbers
(see, e.g., \cite{rogers}). However, for the Bures geometry the
parallels are not so immediate. Consequently, further work is required to
answer (even approximately)
questions of the type: {\sl Given $\varepsilon >0$, what is the
maximal cardinality of a subset of $\cD$ (or $\cS$), every two elements
of which are at least $\varepsilon$ apart in the Bures metric?}

\par
\vskip 3mm {\bf Acknowledgement.} This paper is a part of the
author's Ph.D. dissertation, written under the supervision of Dr.
Stanislaw J. Szarek and Dr. Elisabeth Werner. The author thanks
Dr. Szarek for many valuable discussions and suggestions. The
research has been partially supported by grants from the National
Science Foundation (U.S.A.).  Part of this work was done during
the author's residence at the {\it Workshop in Analysis and
Probability} at Texas A$\&$M University in the summer of 2007. The
author would like to thank the organizers of the workshop and the
Texas A$\&$M Mathematics Department for their  hospitality. The
author thanks the referees for the many helpful suggestions.

\end{document}